\documentclass[a4paper,10pt]{article}
\usepackage{color}
\usepackage{cite}
\usepackage{amsmath,amsthm,amssymb,amsfonts}
\usepackage{eucal}
\usepackage[all]{xy}
\usepackage{color}
\usepackage{fancyhdr}
\usepackage{lastpage}
\usepackage{graphicx}
\usepackage{bm,bbm}
\usepackage{ascmac,paralist}
\usepackage{url}
\usepackage{array,arydshln}
\usepackage{tikz}
\usetikzlibrary{intersections,calc,arrows.meta}

\newtheorem{theorem}{Theorem}[section]

\newtheorem{proposition}[theorem]{Proposition}

\title{Multi-dimensional \\continuous time quantum walks \\related to the birth and death chains}
\author{Yusuke Ide\thanks{Department of Mathematics, College of Humanities and Sciences, Nihon University, Setagaya, Tokyo, 156-8550, Japan, E-mail: ide.yusuke@nihon-u.ac.jp}, 
Norio Konno\thanks{Department of Mathematical Sciences, College of Science and Engineering, Ritsumeikan University, 1-1-1 Noji-higashi, Kusatsu, 525-8577, Japan, E-mail: n-konno@fc.ritsumei.ac.jp}, 
Akihiro Narimatsu\thanks{Center for Mathematical and Data Sciences, The University of Fukuchiyama, Hori 3370, Fukuchiyama, Kyoto, 620-0886, Japan, E-mail: narimatsu-akihiro@fukuchiyama.ac.jp}
}
\date{}
\begin{document}
\maketitle

\begin{abstract}
    In this paper, we consider multi-dimensional birth and death chains and continuous time quantum walks (CTQW) related to them. For CTQW related to our forms of multi-dimensional birth and death chains, we obtain the time scaled independence between multiple dimensions about the transition probability of CTQW. By using this feature, we analyze CTQW on the path graph, which is related to 1-dimensional Ehrenfest model. We also have a random variable which is related to our models and converges to the standard Gaussian distribution. 
\end{abstract}

\section{Introduction}
During recent two decades, the Quantum walks (QWs) has been intensively studied \cite{aharonov, araisan, kempe, kendon, konno3, manouchehri, marq, portugal, venegas} because of the various applicable fields. QWs can be regarded as the quantum counterpart of the classical Random walks (RWs). QWs have several different features from those of RWs \cite{konno1, konno2}. 


On the other hand, the $1$-dimensional Ehrenfest model is widely studied in many fields because this simple model describe many random motions and phenomena \cite{karmc2}. It is known that the Ehrenfest model can be described and analyzed by the tensor product of RWs on the simple edge. In addition, the more general models, called the birth and death chains are also studied in many fields \cite{karmc1, zeifman}. In the previous study \cite{takumikun}, the discrete time quantum walks on the path graph, which are related to the birth and death chains are analyzed. 

In this paper, we consider the multi-dimensional versions of the Ehrenfest model and birth and death chains, which are more general models compared with the Ehrenfest model, and analyze the continuous time Quantum walk (CTQW) related to them. The procedure of the multi-dimensional birth and death chains are to select the dimension and increase or decrease the value of the coordinate of the selected dimension. CTQWs related to the models are regarded as QWs with multi walkers. Then we obtain that the transition probability of the CTQW can be written by the joint probability distribution of independent quantum walkers, whose time is scaled by the select probability of the each walker. 

The rest of this paper is organized as follows. In Section \ref{defi}, we present the definition of our models. Subsection \ref{multiehbd} deals with the 2-dimensional Ehrenfest model as a example, and after that we consider multi-dimensional birth and death chains. In subsection \ref{ctqwdefi}, we define CTQW related to the models given in subsection \ref{multiehbd}. Section \ref{result} is devoted to our main results about CTQW. 
\section{Definition}\label{defi}
In this section, we define the multi-dimensional Ehrenfest model, birth and death chains and CTQW on the path. 

\subsection{The multi-dimensional models}\label{multiehbd}
In this subsection, we focus on the definition the $d$-dimensional Ehrenfest model $(d=1,2,\dots)$.
\subsubsection{$2$-dimensional Ehrenfest model}
At first, we discuss the $2$-dimensional Ehrenfest model as an example. We consider a total of $N$ balls, devided into $2$ kinds, contained in two boxes, $A$ and $B$. Let $N^{(1)}$ be the number of the first kind of balls and $N^{(2)}$ be the number of the second kind of balls in boxes $A$ and $B$, where $N^{(1)}+N^{(2)}=N$. 
We put $X_t^{(i)}\ (i=1,2)$ as the random variable on $\mathbb{Z}$, which denotes the number of $i$-th kind of balls in box $A$ at time $t\in\mathbb{Z}_{\geq 0}$. Then the transition probability of $X_t^{(i)}$ is written as follows:
\begin{align*}
\begin{cases}
\mathbb{P}(X_{t+1}^{(i)}=k\ |\ X_{t}^{(i)}=k)=\frac{N-N^{(i)}}{N}=1-\frac{N^{(i)}}{N}\quad &(k=0,\dots,N^{(i)}),\\
\mathbb{P}(X_{t+1}^{(i)}=k+1\ |\ X_{t}^{(i)}=k)=\frac{N^{(i)}-k}{N}=\frac{N^{(i)}}{N}\bigl(1-\frac{k}{N^{(i)}}\bigr)\quad &(k=0,\dots,N^{(i)}-1),\\
\mathbb{P}(X_{t+1}^{(i)}=k-1\ |\ X_{t}^{(i)}=k)=\frac{k}{N}=\frac{N^{(i)}}{N}\bigl(\frac{k}{N^{(i)}}\bigr)\quad &(k=1,\dots,N^{(i)}),\\
\mathbb{P}(X_{t+1}^{(i)}=j\ |\ X_{t}^{(i)}=k)=0\quad &(\rm{otherwise}).
\end{cases}
\end{align*}
These equations mean that a ball is selected at random and moved to the other box. 
Let $P_N$ be the transition probability matrix of the Ehrenfest model written by 
\begin{align*}
P_N=\frac{N^{(1)}}{N}(P_{N^{(1)}}\otimes I_{N^{(2)}})+\frac{N^{(2)}}{N}(I_{N^{(1)}}\otimes P_{N^{(2)}}),
\end{align*}
where $P_{N^{(i)}}(i=1,2)$ are the transition probability matrices whose $k,l$ elements are $\mathbb{P}(X_{t+1}^{(i)}=l\ |\ X_{t}^{(i)}=k)$. We define the distribution $p_t$ as follows:
\begin{align*}
p_t(x^{(1)},x^{(2)})=\mathbb{P}(X_t^{(1)}=x^{(1)},X_t^{(2)}=x^{(2)}).
\end{align*}
Then we put the operator $M_t:\mathbb{Z}\otimes \mathbb{Z}\to [0,1]$ whose operation is written by
\begin{align*}
M_t=\sum_{0\leq x^{(1)}\leq N^{(1)}}\sum_{0\leq x^{(2)}\leq N^{(2)}}\bigl\{p_t(x^{(1)},x^{(2)})\langle x^{(1)}|\otimes\langle x^{(2)}| \bigr\}.
\end{align*}
We should note that $M_{t+1}=M_tP_N$ and
\begin{align*}
M_t=M_0P_N^t.
\end{align*}
Next, let $\bigl\{\pi^{(i)}(j)\bigr\}\ (i=1,2, j=0,\dots,N^{(i)})$ be the stationary distribution which satisfies the detailed balance condition of the Ehrenfest model. \\
We define the matrix 
\begin{align*}
J_{N^{(i)}}=D_{N^{(i)}}^{1/2}P_{N^{(i)}}D_{N^{(i)}}^{-1/2},
\end{align*}
where $D_{N^{(i)}}=$ diag$\bigl(\pi^{(i)}(0),\dots,\pi^{(i)}(N^{(i)})\bigr)$. Since each of $J_{N^{(i)}}$ is a real symmetric matrix, we have the following spectral decomposition: 
\begin{align}
J_{N^{(i)}}=\sum_{l^{(i)}=0}^{N^{(i)}}\lambda_{l^{(i)}}^{(i)}|{\bf v}_{l^{(i)}}^{(i)}\rangle\langle {\bf v}_{l^{(i)}}^{(i)}|,\label{spedecjni2}
\end{align}
where $\lambda_{l^{(i)}}^{(i)}$ is the $l^{(i)}$-th eigenvalue of $J_{N^{(i)}}$ and 
\begin{align*}
|{\bf v}_{l^{(i)}}^{(i)}\rangle=\begin{bmatrix}
p_{l^{(i)}}^{(i)}(0)\\p_{l^{(i)}}^{(i)}(1)\\ \vdots\\ p_{l^{(i)}}^{(i)}(N^{(i)})
\end{bmatrix}v_{l^{(i)}}^{(i)}=\begin{bmatrix}
1\\p_{l^{(i)}}^{(i)}(1)\\ \vdots\\ p_{l^{(i)}}^{(i)}(N^{(i)})
\end{bmatrix}v_{l^{(i)}}^{(i)}
\end{align*}
is the corresponding eigenvector with $v_{l^{(i)}}^{(i)},\ p_{l^{(i)}}^{(i)}(j)\in\mathbb{R},(j=0,\dots,N^{(i)})$. Noting that $p_{l^{(i)}}^{(i)}$ can be regarded as function of $\lambda_{l^{(i)}}^{(i)}$, we define the measure function $\mu^{(i)}$ by
\begin{align}
\mu^{(i)}(\lambda_{l^{(i)}}^{(i)})=\Bigl|v_{l^{(i)}}^{(i)}\Bigr|^2.\label{omomikansu2}
\end{align}
We put the matrix $J_N$ as follows:
\begin{align*}
J_N=\frac{N^{(1)}}{N}(J_{N^{(1)}}\otimes I_{N^{(2)}})+\frac{N^{(2)}}{N}(I_{N^{(1)}}\otimes J_{N^{(2)}}).
\end{align*}
By Eq.\eqref{spedecjni2}, we obtain
\begin{align*}
J_N\Bigl(|{\bf v}_{l^{(1)}}^{(1)}\rangle \otimes|{\bf v}_{l^{(2)}}^{(2)}\rangle\Bigr)=\Biggl(\frac{N^{(1)}}{N}\lambda_{l^{(1)}}^{(1)}+\frac{N^{(2)}}{N}\lambda_{l^{(2)}}^{(2)} \Biggr)\Bigl(|{\bf v}_{l^{(1)}}^{(1)}\rangle \otimes|{\bf v}_{l^{(2)}}^{(2)}\rangle\Bigr).
\end{align*}
Let $V$ be the orthogonal matrix written by
\begin{align*}
V=\Bigl[
|v_0^{(1)}\rangle\otimes|v_{0}^{(2)}\rangle,\ \ |v_0^{(1)}\rangle\otimes|v_{1}^{(2)}\rangle,\ &\dots\ |v_{0}^{(1)}\rangle\otimes|v_{N^{(2)}}^{(2)}\rangle,\\
|v_1^{(1)}\rangle\otimes|v_{0}^{(2)}\rangle,\ \ |v_1^{(1)}\rangle\otimes|v_{1}^{(2)}\rangle,\ &\dots\ |v_{1}^{(1)}\rangle\otimes|v_{N^{(2)}}^{(2)}\rangle,\\
&\dots\ |v_{N^{(1)}}^{(1)}\rangle\otimes|v_{N^{(2)}}^{(2)}\rangle
\Bigr].
\end{align*}
Since $V$ and $^{\ast}V$ are orthogonal matrices, we have the following another description of $V^{\ast}V=I$:
\begin{align*}
\delta_{{\bf j},{\bf k}}&=\sum_{l^{(1)}=0}^{N^{(1)}}\sum_{l^{(2)}=0}^{N^{(2)}}
\Biggl(\overline{v_{l^{(1)}}^{(1)}p_{l^{(1)}}^{(1)}(j^{(1)})v_{l^{(2)}}^{(2)}p_{l^{(2)}}^{(2)}(j^{(2)})}v_{l^{(1)}}^{(1)}p_{l^{(1)}}^{(1)}(k^{(m)})v_{l^{(2)}}^{(2)}p_{l^{(2)}}^{(2)}(k^{(2)})\Biggr)\\
&=\sum_{l^{(1)}=0}^{N^{(1)}}\sum_{l^{(2)}=0}^{N^{(2)}}\Biggl(\overline{p_{l^{(1)}}^{(1)}(j^{(1)})p_{l^{(2)}}^{(2)}(j^{(2)})}p_{l^{(1)}}^{(1)}(k^{(1)})p_{l^{(2)}}^{(2)}(k^{(2)})|v_{l^{(1)}}^{(1)}|^2|v_{l^{(2)}}^{(2)}|^2 \Biggr)\\
&=\sum_{l^{(1)}=0}^{N^{(1)}}\sum_{l^{(2)}=0}^{N^{(2)}}\Biggl(\overline{p_{l^{(1)}}^{(1)}(j^{(1)})p_{l^{(2)}}^{(2)}(j^{(2)})}p_{l^{(1)}}^{(1)}(k^{(1)})p_{l^{(2)}}^{(2)}(k^{(2)})\mu^{(1)}(\lambda_{l^{(1)}}^{(1)})\mu^{(2)}(\lambda_{l^{(2)}}^{(2)})\Biggr),
\end{align*}
where ${\bf j}=[\ j^{(1)},\ j^{(2)}\ ],\ {\bf k}=[\ k^{(1)},\ k^{(2)}\ ]$ and 
\begin{align*}
\delta_{{\bf j},{\bf k}}=\begin{cases}
1\quad&({\bf j}={\bf k})\\
0\quad&({\bf j}\neq{\bf k}).
\end{cases}
\end{align*}

\subsubsection{Multi-dimensional birth and death chains}
Next we consider more general and higher dimensional models, so-called birth and death chains in the same manner. We consider RW on the $d$-dimensional tensor product $N^{(1)}\otimes\dots\otimes N^{(d)}$ where $N^{(1)}+\dots+N^{(d)}=N$. $i$-th coordinate $(i=1,\dots,d)$ means the number of living $i$-th creature in birth and death chains. Let $q^{(i)}$ be the selection probability of $i$-th dimension among $d$ dimensions and $p^{(i)}$ be the probability fanction that the value of $i$-th coordinate decreases by $1$. We put $X_t^{(i)}\ (i=1,\dots,d)$ as the random variable on $\mathbb{Z}$, which denotes the value of $i$-th coordinate at time $t\in\mathbb{Z}_{\geq 0}$. Then the transition probability of $X_t^{(i)}$ is written as follows:
\begin{align*}
\begin{cases}
\mathbb{P}(X_{t+1}^{(i)}=k\ |\ X_{t}^{(i)}=k)=1-q^{(i)}\quad &(k=0,\dots,N^{(i)}),\\
\mathbb{P}(X_{t+1}^{(i)}=1\ |\ X_{t}^{(i)}=0)=q^{(i)}\times 1,\\
\mathbb{P}(X_{t+1}^{(i)}=k+1\ |\ X_{t}^{(i)}=k)=q^{(i)}(1-p^{(i)}(k))\quad &(k=1,\dots,N^{(i)}-1),\\
\mathbb{P}(X_{t+1}^{(i)}=k-1\ |\ X_{t}^{(i)}=k)=q^{(i)}p^{(i)}(k)\quad &(k=1,\dots,N^{(i)}-1),\\
\mathbb{P}(X_{t+1}^{(i)}=N^{(i)}-1\ |\ X_{t}^{(i)}=N^{(i)})=q^{(i)}\times 1,\\
\mathbb{P}(X_{t+1}^{(i)}=j\ |\ X_{t}^{(i)}=k)=0\quad &(\rm{otherwise}).
\end{cases}
\end{align*}
For simplicity, we assume $0<p^{(i)}(k)<1\ (k=1,\dots,N^{(i)}-1)$. $\mathbb{P}(X_{t+1}^{(i)}=1\ |\ X_{t}^{(i)}=0)=q^{(i)}$ and $\mathbb{P}(X_{t+1}^{(i)}=N^{(i)}-1\ |\ X_{t}^{(i)}=N^{(i)})=q^{(i)}$ means that $p^{(i)}(0)=0,\ p^{(i)}(k)=1$, in other words, we put the reflective boundary conditions for each dimension. 
Let $P_N$ be the transition probability matrix of birth and death chains written by 
\begin{align*}
P_N=q^{(1)}(P_{N^{(1)}}\otimes I_{N^{(2)}}\otimes\cdots\otimes I_{N^{(d)}})+\cdots+q^{(d)}(I_{N^{(1)}}\otimes I_{N^{(2)}}\otimes\cdots\otimes P_{N^{(d)}}),
\end{align*}
where $P_{N^{(i)}}(i=1,\dots,d)$ are the transition probability matrices whose $k,l$ elements are $\mathbb{P}(X_{t+1}^{(i)}=l\ |\ X_{t}^{(i)}=k)$. We define the distribution $p_t$ as follows:
\begin{align*}
p_t(x^{(1)},\dots,x^{(d)})=\mathbb{P}(X_t^{(1)}=x^{(1)},\dots,X_t^{(d)}=x^{(d)}).
\end{align*}
Then we put the operator $M_t:\mathbb{Z}^{\otimes d}\to [0,1]$ whose operation is written by
\begin{align*}
M_t=\sum_{0\leq x^{(1)}\leq N^{(1)}}\cdots\sum_{0\leq x^{(d)}\leq N^{(d)}}\bigl\{p_t(x^{(1)},\dots,x^{(d)})\langle x^{(1)}|\otimes\cdots\otimes\langle x^{(d)}| \bigr\}.
\end{align*}
We should note that $M_{t+1}=M_tP_N$ and
\begin{align*}
M_t=M_0P_N^t.
\end{align*}
Next, let $\bigl\{\pi^{(i)}(j)\bigr\}\ (i=1,\dots,d)$ and $(j=0,\dots,N^{(i)})$ be the stationary distribution which satisfies the detailed balance condition of birth and death chains. \\
We define the matrix 
\begin{align*}
J_{N^{(i)}}=D_{N^{(i)}}^{1/2}P_{N^{(i)}}D_{N^{(i)}}^{-1/2},
\end{align*}
where $D_{N^{(i)}}=$ diag$\bigl(\pi^{(i)}(0),\dots,\pi^{(i)}(N^{(i)})\bigr)$. Since each of $J_{N^{(i)}}$ is a real symmetric matrix, we have the following spectral decomposition: 
\begin{align}
J_{N^{(i)}}=\sum_{l^{(i)}=0}^{N^{(i)}}\lambda_{l^{(i)}}^{(i)}|{\bf v}_{l^{(i)}}^{(i)}\rangle\langle {\bf v}_{l^{(i)}}^{(i)}|,\label{spedecjni}
\end{align}
where $\lambda_{l^{(i)}}^{(i)}$ is the $l^{(i)}$-th eigenvalue of $J_{N^{(i)}}$ and 
\begin{align*}
|{\bf v}_{l^{(i)}}^{(i)}\rangle=\begin{bmatrix}
p_{l^{(i)}}^{(i)}(0)\\p_{l^{(i)}}^{(i)}(1)\\ \vdots\\ p_{l^{(i)}}^{(i)}(N^{(i)})
\end{bmatrix}v_{l^{(i)}}^{(i)}=\begin{bmatrix}
1\\p_{l^{(i)}}^{(i)}(1)\\ \vdots\\ p_{l^{(i)}}^{(i)}(N^{(i)})
\end{bmatrix}v_{l^{(i)}}^{(i)}
\end{align*}
is the corresponding eigenvector with $v_{l^{(i)}}^{(i)},\ p_{l^{(i)}}^{(i)}(j)\in\mathbb{R},(j=0,\dots,N^{(i)})$. Owing to $p_{l^{(i)}}^{(i)}$ can be regarded as function of $\lambda_{l^{(i)}}^{(i)}$, we define the measure function $\mu^{(i)}$ by
\begin{align}
\mu^{(i)}(\lambda_{l^{(i)}}^{(i)})=\Bigl|v_{l^{(i)}}^{(i)}\Bigr|^2.\label{omomikansu}
\end{align}
We put the matrix $J_N$ as follows:
\begin{small}
\begin{align*}
J_N=q^{(1)}(J_{N^{(1)}}\otimes I_{N^{(2)}}\otimes\cdots\otimes I_{N^{(d)}})+\cdots+q^{(d)}(I_{N^{(1)}}\otimes I_{N^{(2)}}\otimes\cdots\otimes J_{N^{(d)}}).
\end{align*}
\end{small}
By Eq.\eqref{spedecjni}, we obtain
\begin{align*}
J_N\Biggl(\bigotimes_{i=1}^{d}|{\bf v}_{l^{(i)}}^{(i)}\rangle \Biggr)=\Biggl(\sum_{i=1}^{d}q^{(i)}\lambda_{l^{(i)}}^{(i)} \Biggr)\Biggl(\bigotimes_{i=1}^{d}|{\bf v}_{l^{(i)}}^{(i)}\rangle \Biggr).
\end{align*}
Let $V_j\ (j=0,\dots,N^{(1)})$ be the orthogonal matrix written by
\begin{align*}
V_j=\begin{bmatrix}
&|{\bf v}_{j}^{(1)}\rangle\otimes|{\bf v}_{0}^{(2)}\rangle\otimes\cdots\otimes|{\bf v}_{0}^{(d-1)}\rangle\otimes |{\bf v}_{0}^{(d)}\rangle,\dots, \\&|{\bf v}_{j}^{(1)}\rangle\otimes|{\bf v}_{0}^{(2)}\rangle\otimes\cdots\otimes|{\bf v}_{0}^{(d-1)}\rangle\otimes |{\bf v}_{N^{(d)}}^{(d)}\rangle,\\
&|{\bf v}_{j}^{(1)}\rangle\otimes|{\bf v}_{0}^{(2)}\rangle\otimes\cdots\otimes|{\bf v}_{1}^{(d-1)}\rangle\otimes |{\bf v}_{0}^{(d)}\rangle,\dots,\\
&|{\bf v}_{j}^{(1)}\rangle\otimes|{\bf v}_{0}^{(2)}\rangle\otimes\cdots\otimes|{\bf v}_{1}^{(d-1)}\rangle\otimes |{\bf v}_{N^{(d)}}^{(d)}\rangle,\dots,\\
&|{\bf v}_{j}^{(1)}\rangle\otimes|{\bf v}_{N^{(2)}}^{(2)}\rangle\otimes\cdots\otimes|{\bf v}_{N^{(d-1)}}^{(d-1)}\rangle\otimes |{\bf v}_{N^{(d)}}^{(d)}\rangle
\end{bmatrix},
\end{align*}
and $V$ be
\begin{align*}
V=\Bigl[\ 
V_0,\ V_1,\dots,\ V_{N^{(1)}}
\ \Bigr].
\end{align*}
Since $V$ and $^TV$ are orthogonal matrices, we have $V^TV=I$ and the following another description:
\begin{align}
\delta_{{\bf j},{\bf k}}
&=\sum_{{\bf l}={\bf 0}}^{[N^{(1)},N^{(2)},\dots,N^{(d)}]}\prod_{m=1}^{d}\overline{\Biggl(v_{l^{(m)}}^{(m)}\Bigl(p_{l^{(m)}}^{(m)}(j^{(m)})\Bigr)\Biggr)}\Biggl(v_{l^{(m)}}^{(m)}\Bigl(p_{l^{(m)}}^{(m)}(k^{(m)})\Bigr)\Biggr)\notag\\
&=\sum_{{\bf l}={\bf 0}}^{[N^{(1)},N^{(2)},\dots,N^{(d)}]}\prod_{m=1}^{d}\overline{\Bigl(p_{l^{(m)}}^{(m)}(j^{(m)})\Bigr)}\Bigl(p_{l^{(m)}}^{(m)}(k^{(m)})\Bigr)\overline{v_{l^{(m)}}^{(m)}}v_{l^{(m)}}^{(m)}\notag\\
&=\sum_{{\bf l}={\bf 0}}^{[N^{(1)},N^{(2)},\dots,N^{(d)}]}\prod_{m=1}^{d}\overline{\Bigl(p_{l^{(m)}}^{(m)}(j^{(m)})\Bigr)}\Bigl(p_{l^{(m)}}^{(m)}(k^{(m)})\Bigr)\mu^{(m)}(\lambda_{l^{(m)}}^{(m)}),\label{orthogo}
\end{align}
where ${\bf j}=[\ j^{(1)},\ j^{(2)},\dots,\  j^{(d)}\ ],\ {\bf k}=[\ k^{(1)},\ k^{(2)},\dots,k^{(d)} \ ]$ and ${\bf l}=[\ l^{(1)},\ l^{(2)},\dots,\ l^{(d)}\ ]$ with 
\begin{align*}
\delta_{{\bf j},{\bf k}}=\begin{cases}
1\quad&({\bf j}={\bf k})\\
0\quad&({\bf j}\neq{\bf k}).
\end{cases}
\end{align*}
Equation \eqref{orthogo} means that $\mu^{(m)}$ can be regarded as the weight function of the orthogonal polynomials $p_{l^{(m)}}^{(m)}$, similar to $2$-dimensional case. 

\subsection{CTQW}\label{ctqwdefi}
This section is devoted to the definition of CTQW related to $J_n$ of birth and death chains.

Let $Y^{(l)}:\mathbb{R}\to\{0,\dots,N^{(l)}\}$ be the random variable about the position of the quantum walker of CTQW, where $l\in\{1,\dots,d\}$. We consider the following state space: 
\begin{align*}
{\mathcal H}={\rm span}\Bigl\{|x^{(1)}\rangle\otimes\dots\otimes|x^{(d)}\rangle\ :\ x^{(j)}\in\{0,\dots,N^{(j)}\},\ j=1,\dots,d\Bigr\}.
\end{align*}
Let $\Psi_t\in{\mathcal H}$ be the state of the quantum walker at time $t\in\mathbb{R}_{\geq 0}$. We define the time evolution operator $U$ of time $\Delta t$ as follows:
\begin{align}
U&(\Delta t)\notag\\
&=e^{i\Delta tJ_N}\notag\\
&=\exp\Biggl[i\Delta t\sum_{{\bf j}={\bf 0}}^{[N^{(1)},N^{(2)},\dots,N^{(d)}]}\Biggl\{\Biggl(\sum_{k=1}^{d}q^{(k)}\lambda_{j^{(k)}}^{(k)}\Biggr)\Biggl(\bigotimes_{l=1}^{d}|{\bf v}_{j^{(l)}}^{(l)}\rangle\Biggr)\Biggl(\bigotimes_{l=1}^{d}\langle {\bf v}_{j^{(l)}}^{(l)}|\Biggr)\Biggr\}\Biggr]\notag\\
&=\sum_{{\bf j}={\bf 0}}^{[N^{(1)},N^{(2)},\dots,N^{(d)}]}\Biggl[\exp\Biggl\{i\Delta t\Biggl(\sum_{k=1}^{d}q^{(k)}\lambda_{j^{(k)}}^{(k)}\Biggr)\Biggr\}\Biggl(\bigotimes_{l=1}^{d}|{\bf v}_{j^{(l)}}^{(l)}\rangle\Biggr)\Biggl(\bigotimes_{l=1}^{d}\langle {\bf v}_{j^{(l)}}^{(l)}|\Biggr)\Biggr],
\label{udeltat}
\end{align}
where ${\bf j}=\{j^{(1)},\dots,j^{(d)}\}$. Then the time evolution of CTQW is written by
\begin{align*}
\Psi_{t+\Delta t}=U(\Delta t)\Psi_{t},
\end{align*}
and the transition probability is given by
\begin{align}
\mathbb{P}({\bf Y}(t+\Delta t)={\bf k}\ |\ {\bf Y}(t)={\bf j})=\Biggl|\Biggl(\bigotimes_{l=1}^{d}\langle k^{(l)}|\Biggr)U(\Delta t)\Biggl(\bigotimes_{l=1}^{d}|j^{(l)}\rangle\Biggr)\Biggr|^2,\label{trproqw}
\end{align}
where ${\bf Y}(t)=\bigl(Y^{(1)}(t),\dots,Y^{(d)}(t)\bigr),$ ${\bf j}=\bigl(j^{(1)},\dots,j^{(d)}\bigr),$ ${\bf k}=\bigl(k^{(1)},\dots,k^{(d)}\bigr),$ $j^{(l)}\in\{0,\dots,N^{(l)}\}$ and $k^{(l)}\in\{0,\dots,N^{(l)}\}$ with $l=1,\dots,d$. We put the transition probability of $l$-th dimension defined by
\begin{align*}
\mathbb{P}^{(l)}\bigl(Y^{(l)}(t+\Delta t)=k^{(l)}\ |\ Y^{(l)}(t)=j^{(l)}\bigr)=\bigl|\langle k^{(l)}|\exp(i\Delta tJ_{N^{(l)}})|j^{(l)}\rangle \bigr|^2.
\end{align*}

\section{Results}\label{result}
This section is devoted to our main results. 
\begin{theorem}\label{theo1}
The transition probability $\mathbb{P}({\bf Y}(t+\Delta t)={\bf k}\ |\ {\bf Y}(t)={\bf j})$ satisfies the following equation:
\begin{small}
\begin{align}
\mathbb{P}({\bf Y}(t+\Delta t)={\bf k}\ |\ {\bf Y}(t)={\bf j})=\prod_{l=1}^{d}\mathbb{P}^{(l)}\Bigl(Y^{(l)}\Bigl(t+q^{(l)}\Delta t\Bigr)=k^{(l)}\ \Bigl|\ Y^{(l)}(t)=j^{(l)}\Bigr).\label{ctqwtrpr}
\end{align}
\end{small}
\end{theorem}
{\it proof of Theorem \ref{theo1}.} By direct calculation, we have
\begin{align}
\mathbb{P}^{(l)}&\bigl(Y^{(l)}(t+\Delta t)=k^{(l)}\ |\ Y^{(l)}(t)=j^{(l)}\bigr)\notag\\
&=\Biggl|\langle k^{(l)}|\sum_{m=0}^{N^{(l)}}\Biggl\{\exp\Bigl(i\Delta t\lambda_{m}^{(l)}\Bigr)|{\bf v}_m^{(l)}\rangle\langle {\bf v}_m^{(l)}|\Biggr\}|j^{(l)}\rangle\Biggr|^2\notag\\
&=\Biggl|\sum_{m=0}^{N^{(l)}}\Biggl\{\exp\Bigl(i\Delta t\lambda_{m}^{(l)}\Bigr)\Bigl(\langle k^{(l)}|{\bf v}_m^{(l)}\rangle\langle {\bf v}_m^{(l)}|j^{(l)}\rangle\Bigr)\Biggr\}\Biggr|^2\notag\\
&=\Biggl|\sum_{m=0}^{N^{(l)}}\Biggl\{\exp\Bigl(i\Delta t\lambda_{m}^{(l)}\Bigr)v_m^{(l)}\Bigl(p_m^{(l)}(k^{(l)})\Bigr)\overline{v_m^{(l)}\Bigl(p_m^{(l)}(j^{(l)})\Bigr)}\Biggr\}\Biggr|^2\notag\\
&=\Biggl|\sum_{m=0}^{N^{(l)}}\Biggl\{\exp\Bigl(i\Delta t\lambda_{m}^{(l)}\Bigr)\Bigl(p_m^{(l)}(k^{(l)})\overline{p_m^{(l)}(j^{(l)})}\Bigr)\mu^{(l)}(\lambda_m^{(l)})\Biggr\}\Biggr|^2.\label{trprlth}
\end{align}
Substituting Eqs.\eqref{udeltat},\eqref{trprlth} into Eq.\eqref{trproqw}, we obtain
\begin{footnotesize}
\begin{align*}
\mathbb{P}&({\bf Y}(t+\Delta t)={\bf k}\ |\ {\bf Y}(t)={\bf j})\notag\\
&=\Biggl|\sum_{{\bf m}={\bf 0}}^{[N^{(1)},N^{(2)},\dots,N^{(d)}]}\Biggl[\exp\Biggl\{i\Delta t\Biggl(\sum_{h=1}^{d}q^{(h)}\lambda_{m^{(h)}}^{(h)}\Biggr)\Biggr\}\Biggl(\prod_{l=1}^{d}\langle k^{(l)}|{\bf v}_{m^{(l)}}^{(l)}\rangle\langle{\bf v}_{m^{(l)}}^{(l)}|j^{(l)}\rangle\Biggr)\Biggr]\Biggr|^2\notag\\
&=\Biggl|\sum_{m^{(1)}=0}^{N^{(1)}}\sum_{{\bf m}=[m^{(1)},0,\dots,0]}^{[m^{(1)},N^{(2)},\dots,N^{(d)}]}\Biggl[\exp\Biggl\{i\Delta t\Biggl(\sum_{h=1}^{d}q^{(h)}\lambda_{m^{(h)}}^{(h)}\Biggr)\Biggr\}\Biggl(\prod_{l=1}^{d}\langle k^{(l)}|{\bf v}_{m^{(l)}}^{(l)}\rangle\langle{\bf v}_{m^{(l)}}^{(l)}|j^{(l)}\rangle\Biggr)\Biggr]\Biggr|^2\notag\\
&=\Biggl|\sum_{m^{(1)}=0}^{N^{(1)}}\Biggl\{\exp\Biggl(iq^{(1)}\Delta t\lambda_{m^{(1)}}^{(1)}\Biggr)\langle k^{(1)}|{\bf v}_{m^{(1)}}^{(1)}\rangle\langle {\bf v}_{m^{(1)}}^{(1)}|j^{(1)}\rangle\Biggr\}\Biggr|^2\notag\\
&\quad\times\Biggl|\sum_{{\bf m}=[\ast,0,\dots,0]}^{[\ast,N^{(2)},\dots,N^{(d)}]}\Biggl[\exp\Biggl\{i\Delta t\Biggl(\sum_{h=2}^{d}q^{(h)}\lambda_{m^{(h)}}^{(h)}\Biggr)\Biggr\}\Biggl(\prod_{l=2}^{d}\langle k^{(l)}|{\bf v}_{m^{(l)}}^{(l)}\rangle\langle{\bf v}_{m^{(l)}}^{(l)}|j^{(l)}\rangle\Biggr)\Biggr]\Biggr|^2
\notag\\
&=\mathbb{P}^{(1)}\Bigl(Y^{(1)}\Bigl(t+q^{(1)}\Delta t\Bigr)=k^{(1)}\ \Bigl|\ Y^{(1)}(t)=j^{(1)}\Bigr)\notag\\
&\quad\times\Biggl|\sum_{{\bf m}=[\ast,0,\dots,0]}^{[\ast,N^{(2)},\dots,N^{(d)}]}\Biggl[\exp\Biggl\{i\Delta t\Biggl(\sum_{h=2}^{d}q^{(h)}\lambda_{m^{(h)}}^{(h)}\Biggr)\Biggr\}\Biggl(\prod_{l=2}^{d}\langle k^{(l)}|{\bf v}_{m^{(l)}}^{(l)}\rangle\langle{\bf v}_{m^{(l)}}^{(l)}|j^{(l)}\rangle\Biggr)\Biggr]\Biggr|^2
\notag\\
&=\prod_{l=1}^{d}\mathbb{P}^{(l)}\Bigl(Y^{(l)}\Bigl(t+q^{(l)}\Delta t\Bigr)=k^{(l)}\ \Bigl|\ Y^{(l)}(t)=j^{(l)}\Bigr).
\end{align*}
\end{footnotesize}
This equation means that the transition probability of multi-dimensional CTQW can be calculated by the product of the transition probability of $1$-dimensional CTQW.\hfill$\square$

From now on, we assume $N^{(l)}=N/d\ (l=1,\dots,d)$ and $q^{(l)}=1/d\ (l=1,\dots,d)$ . Let ${\bf j}=[j^{(1)},\dots,j^{(d)}]$ be the initial position of the quantum walker and 
\begin{align*}
{\bf Y}_{\bf j}(t)=[Y^{(1)}_{j^{(1)}}(t),\dots,Y^{(d)}_{j^{(d)}}(t)]
\end{align*}
be the position at time $t\in\mathbb{R}$. Each $Y_{j^{(l)}}^{(l)}\in\{0,\dots,N/d\}$ can be regarded as the random variable. By Eq.\eqref{ctqwtrpr}, the transition probability is written as follows:
\begin{align}
\mathbb{P}({\bf Y}_{\bf j}(t)={\bf k})&=\prod_{l=1}^{d}\mathbb{P}^{(l)}\Bigl(Y_{j^{(l)}}^{(l)}\Bigl(q^{(l)}t\Bigr)=k^{(l)}\Bigr)\notag\\
&=\prod_{l=1}^{d}\mathbb{P}^{(l)}\Bigl(Y_{j^{(l)}}^{(l)}\Bigl(\frac{t}{d}\Bigr)=k^{(l)}\Bigr).\label{ctqwiid}
\end{align}
When $t$ of left-hand side of Eq.\eqref{ctqwiid} is scaled to $t/d$, this equation means that $Y_{j^{(l)}}^{(l)},(l=1,\dots,d)$ are independent. When we assume $N^{(l)}=1\ (l=1,\dots,d=N)$, we can regard CTQW on the path as the tensor product by $N$-th power of CTQW on the edge. Thus we have the following proposition. 
\begin{proposition}\label{prop1}
The random variable about the position of CTQW related to the $1$-dimensional Ehrenfest model $Y_0=Y^{(1)}_0+Y^{(2)}_0+\dots,+Y^{(d)}_0$ obeys the binomial distribution $B(d, \sin ^2\frac{t}{d})$.
\end{proposition}
{\it proof of Proposition \ref{prop1}.} For each $l=0,\dots,d$, we have
\begin{align*}
J_{N^{(l)}}=J_{1}=\begin{bmatrix}
0&1\\1&0
\end{bmatrix}.
\end{align*}
Then we obtain
\begin{align*}
e^{itJ_{N^{(l)}}}&=\sum_{k=0}^{\infty}\frac{(it)^k}{k!}J_{N^{(l)}}^k\\
&=\sum_{k=0}^{\infty}\frac{(it)^{2k}}{(2k)!}\begin{bmatrix}
    1&0\\0&1
\end{bmatrix}+\sum_{k=0}^{\infty}\frac{(it)^{2k+1}}{(2k+1)!}\begin{bmatrix}
    0&1\\1&0
\end{bmatrix}\\
&=\begin{bmatrix}
    \cos t&i\sin t\\
    i\sin t&\cos t
\end{bmatrix}.
\end{align*}
Then we get $\mathbb{P}^{(l)}\Bigl(Y_{0}^{(l)}(t)=1\Bigr)=\sin^2 t$. By Eq.\eqref{ctqwiid}, we have
\begin{align*}
\mathbb{P}\left(Y_0(t)=\sum_{l=1}^{d}j^{(l)}\right)&=\mathbb{P}({\bf Y}_{\bf 0}(t)={\bf j})\\
&=\prod_{l=1}^{d}\mathbb{P}^{(l)}\Bigl(Y_{0}^{(l)}\Bigl(\frac{t}{d}\Bigr)=j^{(l)}\Bigr),
\end{align*}
where ${\bf j}=[j^{(1)},\dots,j^{(d)}]$, $j^{(l)}\in\{0,1\}\ (l=1,\dots,d)$. Thus we obtain the conclusion. 
\hfill $\square$

This is another easier calculation method for the analysis of the $1$-dimensional CTQW related to Ehrenfest model, obtained in \cite{konno4}. 

Next we consider the random variable $Z_{\bf j}^{(d)}$ of position at time $T\in\mathbb{R}$, defined by
\begin{align*}
Z_{\bf j}^{(d)}(T)=\frac{1}{d}\bigl\{Y_{j^{(1)}}^{(1)}(T)+\dots+Y_{j^{(d)}}^{(d)}(T) \bigr\}.
\end{align*}
\begin{theorem}\label{theo2}
For any constant $T\geq 0$, we have
\begin{align}
\frac{Z_{\bf j}^{(d)}(T)-\mathbb{E}\bigl(Y^{(1)}_{j^{(1)}}(T)\bigr)}{\sqrt{\frac{V\bigl(Y^{(1)}_{j^{(1)}}(T)\bigr)}{d}}}\Rightarrow N(0,1)\quad (d\to \infty),\label{theoeq}
\end{align}
where $\Rightarrow$ stands for the weak convergence and $N(0,1)$ is the standard Gaussian distribution.
\end{theorem}
{\it proof of Theorem \ref{theo2}.}
Noting that $Y_{j^{(1)}}^{(1)}(T),\dots,Y_{j^{(d)}}^{(d)}(T)$ are independent and identically distributed. Since $T$ is a constant, variance of $Y_{j^{(1)}}^{(1)}(T),\dots,Y_{j^{(d)}}^{(d)}(T)$ are finite. Then we have Eq. \eqref{theoeq} by central limit theorem.
\hfill $\square$

\section{Summary}
In this paper, we considered multi-dimensional birth and death chains and CTQW related to them. We obtained the time scaled independence between multiple dimensions about the transition probability of CTQW. After that we limited our discussion to the symmetric multi-dimensional Ehrenfest model case. We got the different proof from the previous study to show that the random variable about the position of CTQW related to the 1-dimensional Ehrenfest model obeys the binomial distribution. We also had a random variable which is related to our models and converges to the standard Gaussian distribution. To consider CTQW on more general graph and analyze by using the dividing method of proposition \ref{prop1} is one of the interesting future problem.


\begin{thebibliography}{}
\bibitem{aharonov}
D. Aharonov, A. Ambainis, J. Kempe, U.V. Vazirani, Quantum walks on graphs, In: Proceedings of
the 33rd Annual ACM Symposium on Theory of Computing (STOC ’01), pp. 50–59, 2001.

\bibitem{araisan}
T. Arai, CL. Ho, Y. Ide and N. Konno, Periodicity for space-inhomogeneous quantum walks on the cycle, {\it Yokohama Math. J.} {\bf 62}, 39-50, 2016.

\bibitem{takumikun}
CL. Ho, Y. Ide, N. Konno, E. Segawa and K. Takumi, A Spectral Analysis of Discrete-Time Quantum walks Related to the Birth and Death Chains, {\it J. Stat. Phys.}, {\bf 171}, pp. 207–219, 2018.

\bibitem{karmc1}
S. Karlin and J. L. McGregor, The Differential Equations of Birth-and-Death Processes, and the Stieltjes Moment Problem, {\it Trans. Amer. Math. Soc.}, {\bf 85} (2), pp. 489-546, 1957.

\bibitem{karmc2}
S. Karlin and J. McGregor, Ehrenfest urn models, {\it J. Appl. Probab.}, {\bf 2} (2), pp. 352-376, 1965.

\bibitem{kempe}
J. Kempe, Quantum random walks - an introductory overview, {\it Contemp. Phys.,} {\bf 44}, pp. 307-327, 2003.

\bibitem{kendon}
V. Kendon, Decoherence in quantum walks - a review. Math. Struct, {\it Comp. Sci.,} {\bf 17}, pp. 1169-1220, 2007.
\bibitem{konno1}
N. Konno, Quantum random walks in one dimension, {\it Quant. Inf. Process.,} {\bf 1}, pp. 345-354, 2002.
\bibitem{konno2}
N. Konno, A new type of limit theorems for the one-dimensional quantum random walk, {\it J. Math. Soc. Jpn.,}
{\bf 57}, pp. 1179-1195, 2005.

\bibitem{konno4}
N. Konno, Continuous-time quantum walks on ultrametric spaces, {\it Int. J. Quantum Inf.,} {\bf 04} (06), pp. 1023-1035, 2006.

\bibitem{konno3}
N. Konno, Quantum walks, In: Quantum Potential Theory, U. Franz and M. Sch\"{u}rmann, Eds., Lecture Notes in Mathematics:, {\bf 1954}, pp. 309-452, {\it Springer-Verlag}, Heidelberg, 2008.

\bibitem{manouchehri}
K. Manouchehri and J.B. Wang, Physical Implementation of Quantum Walks, {\it Springer-Verlag}, Heidelberg, 2013.

\bibitem{marq}
F.L. Marquezino, R. Portugal, G. Abal and R. Donangelo, Mixing times in quantumwalks on the hypercube,
{\it Phys. Rev.} {\bf A} 77, 042312, 2008.

\bibitem{portugal}
R. Portugal, Quantum walks and search algorithms, {\it Springer-Verlag}, New York, 2013.

\bibitem{venegas}
S.E. Venegas-Andraca, Quantum walks: A comprehensive review, {\it Quantum Inf. Process.,} {\bf 11}, pp. 1015-1106, 2012.

\bibitem{zeifman}
A. Zeifman, S. Leorato, E. Orsingher, Ya. Satin and G. Shilova, Some universal limits for nonhomogeneous birth and death processes, {\it Queueing Syst.}, {\bf 52}, pp. 139–151, 2006.
\end{thebibliography}
\end{document}